\documentclass[aps,twocolumn,showpacs,amsmath,amssymb,pre,superscriptaddress,floatfix]{revtex4-1}

\usepackage{graphicx}
\usepackage{dcolumn}
\usepackage{bm}

\begin{document}

\title{Phase Space Interpretation of Exponential Fermi Acceleration}

\date{\today}

\pacs{05.45.a,05.45.Ac,05.45.Pq}

\author{Benno Liebchen}
\email[]{Benno.Liebchen@physnet.uni-hamburg.de} 
\affiliation{Zentrum f\"ur Optische Quantentechnologien, Universit\"at Hamburg,
Luruper Chaussee 149, 22761 Hamburg, Germany}
\author{Robert B\"uchner}
\affiliation{Zentrum f\"ur Optische Quantentechnologien, Universit\"at Hamburg,
Luruper Chaussee 149, 22761 Hamburg, Germany}
\author{Christoph Petri}
\affiliation{Zentrum f\"ur Optische Quantentechnologien, Universit\"at Hamburg,
Luruper Chaussee 149, 22761 Hamburg, Germany}
\author{Fotis K. Diakonos}
\affiliation{Department of Physics, University of Athens, GR-15771 Athens,
Greece}
\author{Florian Lenz}
\affiliation{Zentrum f\"ur Optische Quantentechnologien, Universit\"at Hamburg,
Luruper Chaussee 149, 22761 Hamburg, Germany}
\author{Peter Schmelcher}
\email[]{Peter.Schmelcher@physnet.uni-hamburg.de} 
\affiliation{Zentrum f\"ur Optische Quantentechnologien, Universit\"at Hamburg,
Luruper Chaussee 149, 22761 Hamburg, Germany}

\begin{abstract}
Recently, the occurrence of exponential Fermi acceleration has been reported  
in a rectangular billiard with an oscillating bar inside [K. Shah, D. Turaev, and V. Rom-Kedar, Phys. Rev. E {\bf 81}, 056205 (2010)]. 
In the present work, we analyze the underlying physical mechanism and show that the phenomenon can be understood as a sequence of highly correlated motions, consisting of alternating phases 
of free propagation and motion along the invariant spanning curves of the well-known one-dimensional Fermi-Ulam model.
The key mechanism for the occurrence of exponential Fermi acceleration
can be captured in a random walk model in velocity space with step width proportional to the velocity itself. The model reproduces
the occurrence of exponential Fermi acceleration and provides a good ab initio prediction of the value of the growth rate including its full parameter-dependency.
Our analysis clearly points out the requirements for exponential Fermi acceleration, thereby opening the perspective of finding other systems 
exhibiting this unusual behaviour.
\end{abstract}

\maketitle

\section{Introduction}

Recently, the investigation of Fermi acceleration (FA) in two-dimensional (2D)
time-dependent billiards has attracted a lot of attention  \cite{Loskutov:1999,Loskutov:2002,Carvalho:2006,Kamphorst:2007,Lenz:2008, Bunimovich:2010,Gelfreich:2011}. Fermi acceleration is
the unbounded energy gain of particles exposed to driving forces and was first
proposed in 1949 by Enrico Fermi \cite{Fermi:1949} to explain the high energies
of cosmic radiation (for a review of FA see Ref. 
\cite{Blandford1987}). He suggested that charged particles repeatedly interact
with time-dependent magnetic fields (originating either from shock waves of
supernovae or from magnetized interstellar clouds) in such a way that on average
they gain energy. Nowadays, FA is investigated in a variety of  systems
belonging to different areas of physics, such as astrophysics
\cite{Veltri2004,Honda2002,Malkov1998}, plasma physics
\cite{Michalek1999,Milovanov2001}, atom optics
\cite{Saif1998,Steane1995} and has even been used for the interpretation of
experimental results in atomic physics \cite{Lanzano1999}.

The one-dimensional (1D) prototype system allowing the investigation of FA is
the so-called Fermi-Ulam model (FUM) \cite{Lichtenberg:1992}, which consists of
non-interacting particles moving between one fixed and one oscillating wall. The
FUM and its variants have been the
subject of extensive theoretical (see Ref. \cite{Lichtenberg:1992, Karlis:2006}
and references therein) and experimental \cite{Kowalik:1988,Celaschi:1987}
studies. In the FUM, the existence of FA depends exclusively on the
driving-law of the oscillating wall: As long as the driving law is
sufficiently smooth, there is no unlimited energy growth due to the existence
of invariant spanning curves \cite{Lichtenberg:1992}. In particular, this means
that harmonic driving laws do not lead to FA in the FUM. 

In 2D time-dependent billiards, already a smooth driving law may lead to FA. For
example, the existence of FA was shown for
a harmonically oscillating stadium-like billiard  \cite{Loskutov:1999,Loskutov:2002},
in the driven eccentric annular billiard \cite{Carvalho:2006}, in an oval billiard \cite{Kamphorst:2007} and in
the time-dependent elliptical billiard \cite{Lenz:2008, Lenz2009, Lenz2010}. On the other
hand, the breathing concentric
annular \cite{Carvalho:2006} and the circular billiard \cite{Kamphorst:1999} do
not exhibit FA. The ensemble-averaged energy $E(t)$ in all the 2D time-dependent billiards that do show FA, grows as a function of time according to a power law, i.e. $E(t) \sim t^{d}$, with some exponent $d$. 
For such power laws, it is known that FA is not structurally stable in the sense that any finite amount of dissipation will destroy it \cite{Bunimovich:2010, Leonel:2010, Oliveira:2011, Petri:2010}, 
independent on whether dissipation is introduced via inelastic collisions or via drag forces (e.g. Stokes' friction).

Thus, a natural question to ask is whether there are certain time-dependent billiards that show a somewhat `faster' acceleration of the energy, in particular whether there is e.g. exponential acceleration possible. The first hint that such a fast acceleration process is possible is given in Ref. \cite{Gelfreich:2008}, where the authors prove the existence of single orbits with exponential energy growth under certain conditions. 
In Ref. \cite{Shah:2010} it is shown that the energy of a whole ensemble of particles grows exponentially in a rectangular billiard with an oscillating bar inside. 
This result is generalized in Ref. \cite{Gelfreich:2011}, where the authors show by means of an analysis of the Anosov-Kasuga invariant that in  special classes of billiard systems the ensemble averaged energy accelerates exponentially. 

While in Refs. \cite{Gelfreich:2008, Shah:2010, Gelfreich:2011} a mathematical analysis of exponential FA is provided, the aim of the present work is to investigate the physical mechanism leading to exponential FA in the setup proposed in Ref. \cite{Shah:2010}. This setup consists of an oscillating bar inside a rectangular billiard, where the bar is aligned parallel to the long side of the rectangle. 
This can be interpreted as particles moving alternately in a FUM and in a static rectangular billiard. Since neither the static rectangular billiard nor the FUM alone even show FA,  
we want to clarify from a physical point of view how the combination of the both leads to exponential FA, i.e. what are the microscopic processes that cause the astonishingly fast acceleration. 
To this end, we will show that in the high velocity regime, the temporal movement of the particles on invariant curves of the FUM can be 
modeled by a suitable random walk with step sizes being proportional to the velocity itself. 
This random walk model shows exponential acceleration. Furthermore, the corresponding parameters of the random walk can be extracted from the underlying FUM, 
even enabling an alternative (compared with the one given in Ref. \cite{Shah:2010}) prediction of the exponential acceleration rate without any free parameters.     

The work is structured as follows: In section \ref{ch:setup} we introduce the setup and show the results of our numerical simulations. How the microscopic dynamics of single trajectories can be interpreted as piecewise motion along invariant spanning curves of an appropriate Fermi-Ulam model is shown in section \ref{ch:fum}. 
With this picture in mind, we construct a random walk model in section \ref{ch:rwm} that generalizes the considered model incorporating the basic characteristics of the underlying physical mechanism.
Finally, a short summary and outlook is given in section \ref{ch:conclusion}.

\section{Setup and Results}\label{ch:setup}
\begin{figure}
\includegraphics[width=8.6cm]{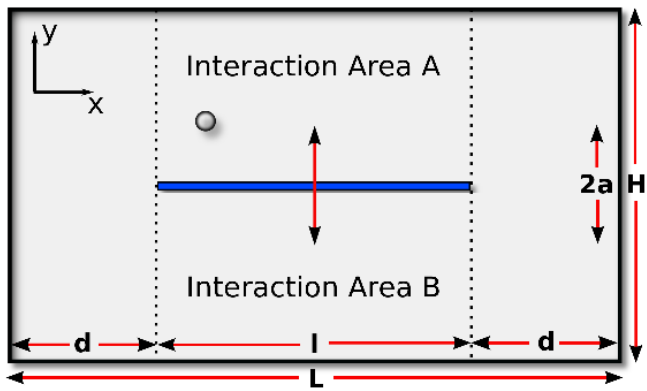}
\caption{(Color online) Setup: Rectangular billiard of height $H$ and length $L$ with an oscillating bar of length $l$ placed in the center, parallel to the $x$-axis (here $H=2$, $L=4$, $l=2$). 
The two areas above and below the bar are the interactions areas (with the moving bar) A and B, respectively. \label{fig:fig1}}
\label{setup}
\end{figure}
\begin{figure}
\includegraphics[width=8.6cm]{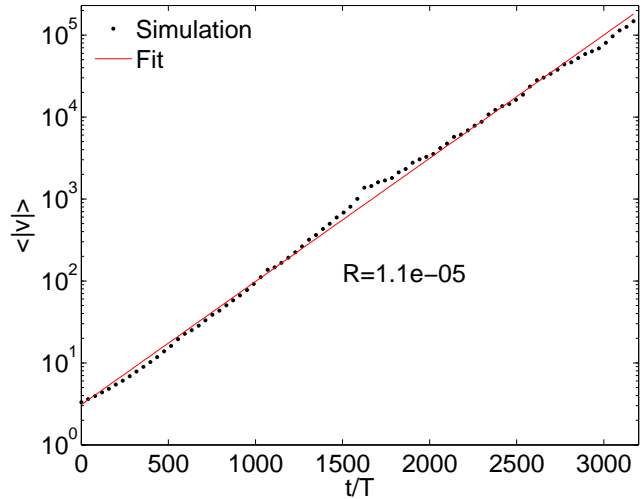}
\caption{(Color online) Semi-logarithmic plot of the time-evolution of the ensemble averaged modulus of the velocity $\langle |v|\rangle(t)$. The velocity grows exponentially, $\langle |v|\rangle (t) \sim e^{R t}$, with a growth rate of $R=1.1\times 10^5$ ($a=0.1$, $\omega=0.02$). \label{fig:fig2}}
\label{grate}
\end{figure}

The investigated setup is shown in Fig.~\ref{fig:fig1}. It consists of a rectangular billiard of length $L$ and height $H$ with an oscillating bar inside. The rectangular billiard without the bar is integrable, since upon collisions with the billiard boundary, just the sign of the corresponding component of the velocity $\bm v$ is reversed, 
i.e. in particular $|v_x|$ and $|v_y|$ are preserved. 
Now, a bar of length $l$ is placed in the middle of the billiard, parallel to the longer side ($x$-direction) of the billiard. 
Here, we assume an harmonic oscillation law: $y_b=a \cos(\omega t)$,
with $a$ and $\omega$ being the driving amplitude and frequency, respectively. Since the oscillating bar transfers momentum in $y$-direction only, $|v_x|$ is preserved. 
As an initial ensemble, we take $N=10^4$ classical, non-interacting particles with a fixed velocity $v_x=0.16$; and $v_y$ is randomly chosen in the interval $[0,40v_x]$ (we use $H=2$, $L=4$, $l=2$ for the simulations). We iterate these particles by numerically solving the corresponding discrete mapping, i.e. by calculating the successive collisions with the billiard boundary, which consists of the rectangle and the bar. 
The main computational effort is to determine the time of the next collision with the oscillating bar, where the smallest root of an implicit equation (with possibly many roots) has to be found, see e.g. Refs.~\cite{Koch:2008,Lenz:2007}. 
The main quantity of interest is the time-evolution of the ensemble averaged modulus of the velocity in $y$-direction ($v_x=const$; $v_y \equiv v$), which is given by 
\begin{equation}
 \langle |v|\rangle(t) = \frac{1}{N}\sum_{i=1}^N |v_i(t)|,
\end{equation}
where $v_i(t)$ is velocity of the $i$-th particle at time $t$. The results of the simulations are shown in Fig.~\ref{fig:fig2} for 3000 oscillations of the bar on a semilogarithmic scale. 
The ensemble averaged velocity clearly grows exponentially $\langle |v|\rangle(t)\sim \exp(R t)$ as reported in Ref.~\cite{Shah:2010}, with a growth rate of $R \approx 1.1\times 10^{-5}$. 
Let us now develop a physical picture of this acceleration process, answering the question how we can link the microscopic dynamics of single trajectories to the appearance of exponential Fermi acceleration. 

\section{Connection with the Fermi-Ulam model}\label{ch:fum}
Since the $v_x$ component of a particle's velocity stays constant, see Fig.~\ref{fig:fig1}, as long as the particle is in one of the interaction areas, the dynamics in $y$ direction corresponds exactly to the one of a particle moving 
in an one-dimensional (1D) Fermi-Ulam model (FUM), where the 
distance between the equilibrium position of the moving wall and the static wall is given by $h=H/2$. 
The time $t_I$ the particle spends inside the interaction area is simply given by $t_I=l/v_x$. 
We define the time $t_F$ the particle spends in the FUM by each passing of the interaction area (A or B, see Fig.~\ref{setup}) as the time difference between the first and the 
last collision with the oscillating bar while the particle is in the interaction area. The motivation for this definition 
is that only collisions with the oscillating bar change $v_y$, i.e. we want to keep track of the time in which a certain change in the velocity takes place. 
The times $t_I$ and $t_F$ are not identical, since once a particle enters the interaction area, a certain amount of time will elapse, before it collides with the bar. 
However, for high velocities $\bm v$, which means that $v_y$ is large since $v_x=const.$, $t_F$ converges towards $t_I$. By `high' we mean 
that $\bm v$ is large compared to the maximal velocity of the bar, i.e. for $|\bm v| \gg \omega a_0$ (which implies $|\bm v|\approx |v_y|$) we obtain $t_F\approx t_I = l/v_x$.   
\begin{figure}
\includegraphics[width=8.6cm]{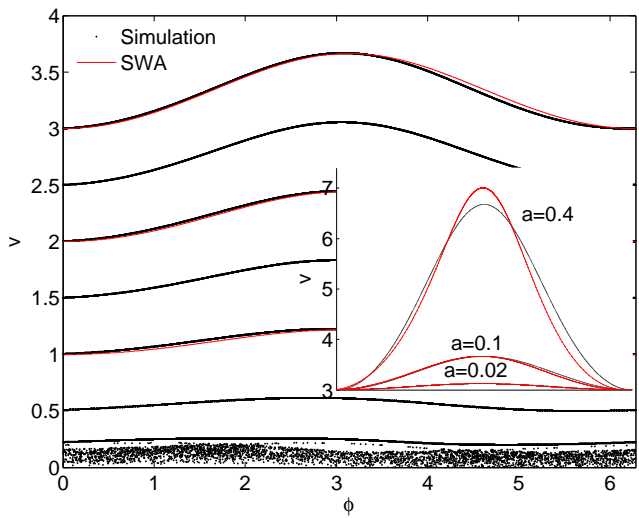}
\caption{(Color online) Phase space of the one-dimensional Fermi-Ulam model. 
For low $v$ there is a chaotic sea, whereas for high velocities there are invariant spanning curves. 
The difference between the maximum and the minimum of such a curve grows with increasing $v$.  
The red lines show the analytic results for the invariant spanning curves based on the static wall approximation (SWA). 
In the inset, a single invariant spanning curve is shown for different driving amplitudes $a$. For large $a$, the SWA deviates significantly from the exact result. \label{fig:fum}}
\label{fumiscs}
\end{figure}

Since the dynamics of particles can be described for some time spans $t_F$ as a 1D FUM, it is convenient to summarize some of the properties of the FUM (for a more detailed description see Ref.~\cite{Lichtenberg:1992} and references therein). The phase space of the FUM is shown in Fig.~\ref{fig:fum}. 
For low velocities there is a large chaotic sea containing many regular islands (this regime is shrinked to a narrow band $0 < v \lesssim 0.3$ in Fig.~\ref{fumiscs}). Above the first invariant spanning curve (FISC) with velocity $v_c$, the motion becomes more and more regular, 
until for $v\gg v_c$ there are exclusively invariant spanning curves corresponding to a synchronized motion between the oscillating wall and the particles. Due to these invariant curves, 
there is no diffusion in momentum space and the FUM with a harmonic oscillation of the wall does not show Fermi acceleration.  

The invariant spanning curves $v^{\text{isc}}(\phi)$ are not just straight lines, but show a characteristic shape (see Fig.~\ref{fig:fum}). 
There are infinitely many of them, which can be labeled by the velocity $v^{\text{isc}}(\phi=0)$ and are parametrized as $v^{\text{isc}}=v^{\text{isc}}(\phi,\tilde v)$.
The minimum  of these curves is always at $\phi=0$ and the maximum at $\phi=\pi/\omega$. 
The difference $\triangle v^{\text{isc}}= v^{\text{isc}}(\pi,\tilde v)-v^{\text{isc}}(0,\tilde v)$ grows linearly with increasing $\tilde v$, i.e. $\triangle v^{\text{isc}}\sim \tilde v$. 
For high velocities ($v\gg v_c$) this can be rigorously shown within the so-called static wall approximation (SWA) \cite{Lichtenberg:1992,Karlis:2006}, which assumes that the oscillating 
bar is fixed in coordinate space but transfers momentum as if it would be moving. 
The distance between two collisions is then simply $2h$ and the time between two collisions is $\triangle t = 2h/v$, where $v$ is
the velocity after the preceding wall collision. After a collision with the (only in momentum space moving) wall, the velocity $v_1$ of the particle is
\begin{equation}
v_1 = v^{\rm isc}(t,\tilde v) - 2 v_w(t),
\end{equation}
where $v_w(t)= \dot y_b(t)=-a \omega \sin(\omega t)$ is the velocity of the wall. 
Since for high velocity the particle moves on an invariant curve we set $v_1=v^{\rm isc}(t+\Delta t,\tilde v)$ and get
\begin{equation}
v^{\rm isc}(t+\Delta t,\tilde v) = v^{\rm isc}(t,\tilde v) - 2 v_w(t).
\end{equation}
Expanding this equation into a Taylor series up to first order and applying the continuous limit $\triangle t \rightarrow 0$, we obtain $2h \dot v^{\rm isc}(t,\tilde v)/v^{\rm isc}(t,\tilde v) = -2v_w(t)$. Integration yields
\begin{equation}\label{eq:isc_swa}
v^{\rm isc}(\phi,\tilde v)=\tilde v \;{\rm e}^{(a/h) \cdot [1-\cos \phi]}\;\quad \phi\in [0,2\pi)
\end{equation}
Therein, $\phi$ is the phase of the wall oscillation, $\phi=\omega t \mod 2\pi$. Obviously, the difference $\triangle v^{\rm isc}$ between the velocity maximum and minimum of an invariant spanning curve is 
\begin{equation}
 \triangle v^{\rm isc}=v^{\rm isc}(\pi,\tilde v)-v^{\rm isc}(0,\tilde v)=\tilde v \cdot \left( {\rm e}^{2a/h}-1\right)\sim \tilde v
\end{equation}
and thus proportional to $\tilde v$. The $v^{\rm isc}(\phi,\tilde v)$ of Eq.~\eqref{eq:isc_swa} are shown in Fig.~\ref{fig:fum} as red lines. We obtain a very good agreement with the exact results from the numerical simulations. 
However, the inset shows a single invariant spanning curve for different driving amplitudes $a$,  and for large $a$, the SWA (Eq.~\ref{eq:isc_swa}) deviates significantly from the exact result, 
i.e. the SWA is valid for small driving amplitudes only.
 
\begin{figure}
\includegraphics[width=8.6cm]{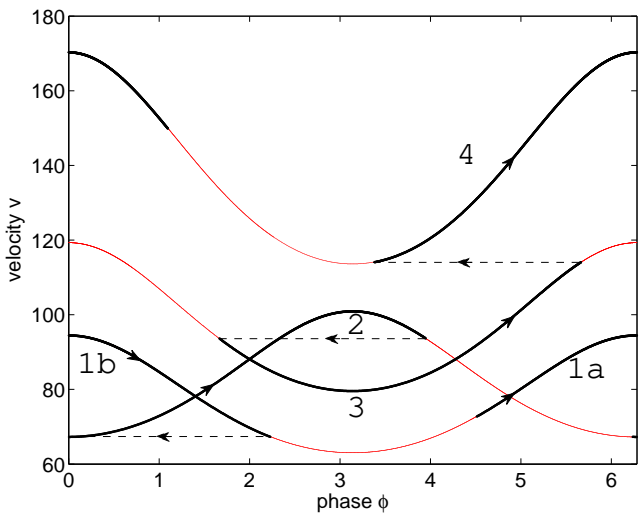}
\caption{(Color online) A typical trajectory in the investigated setup (see Fig.~\ref{fig:fig1}). The particle enters four times the interaction area (curves 1,3,4 area B, curve 2 area A). For a fixed time $t_F$ (or phase $\triangle \phi= \omega t_F \mod 2\pi$) it moves along the invariant spanning curves of the corresponding Fermi-Ulam model (thin red lines), see also Fig.~\ref{fig:fum}. It then leaves the interaction area with a certain velocity before it re-enters area A or B with the same velocity but now with a different phase.   \label{fig:fig3}}
\end{figure}
Based on the above phase space description of the FUM we can investigate the microscopic
dynamics of a typical trajectory moving inside the oscillating bar billiard. 
Since we are ultimately interested in the acceleration process, we can assume that the particles are fast compared to the motion of the bar, i.e. again $v\gg a_0 \omega$. 

This means that the velocity is also much larger than the velocity $v_c$ of the FISC. 
A particle enters the interaction area, let's say above the bar (interaction area A), at time $t_1$ with a certain velocity $v_1=|\bm v_1|$ and 
spends the time $t_I\approx t_F$ in it. Since it is fast, it collides many times during the time $t_F$ with the bar, thus 
moving along the corresponding invariant spanning curve of the FUM and leaving the interaction area at time $t_2=t_1+t_F$ with velocity $v_2$. 
Now the particle propagates in the free part of the rectangular billiard (i.e. the part where no collisions with the bar take place) for a time $t_b=2d/v_x$. 
During this time, the modulus of the velocity does not change and thus the particle re-enters at time $t_3=t_2+t_b$ the interaction area with velocity  $v_2$. 
The particle can enter either the interaction area A or B, i.e. above or below the bar depending on the exact dynamics. However, for high velocities, in a good approximation 
this is a random process, as argued in Ref~\cite{Shah:2010}, i.e. the particle will be injected with probability one half above and with probability one half below the bar. 
If the particle is injected in part A, it re-enters the same FUM as described above, now at the phase $\phi=\omega t_3\mod 2\pi$. 
If it is injected in part B we have to add a phase shift of $\pi$, i.e. $\phi=(\omega t_3+\pi)\mod 2\pi$, since the two FUM's (above and below the bar) 
can be transformed into each other simply by shifting the phase by $\pi$. Again, the particle moves for a time $t_F$ along an invariant curve and the whole 
described process - random injection in the FUM, leaving the FUM, re-entering etc. - starts over again. Exemplary, this process is shown for a typical 
trajectory in Fig.~\ref{fig:fig3}. The thick black lines show the motion along the invariant spanning curves and thus the evolution of the velocity of the particle 
for four such injections into the interaction area. The thin red lines show the invariant spanning curves of the corresponding FUM's. We see perfect agreement, supporting our assumption
that the particle moves for some time $t_F$ along such invariant curves. Note that in the above terminology, 
the curve 2 corresponds to the FUM of the interaction area A, whereas the curves 1,3,4 correspond to the phase shifted FUM of the interaction area B.

Let us describe the above process in a more quantitative way. 
The particle enters the interaction area at $t_1$ with a high velocity $v_1\gg a_0\omega$ and $v_1\gg v_x$. 
The corresponding invariant spanning curve on which the particle will move for the time $t_F$ can be calculated as follows: 
The entry phase $\phi_1$ is given by $\phi_1=\omega t_1\mod 2\pi$; by setting $v_1=v^{\rm isc}(\phi_1,\tilde v)=\tilde v{\rm e}^{(a_0/h) \cdot [1-\cos \phi_1]}$ 
we obtain $\tilde v = v_1 {\rm e}^{(-a_0/h) \cdot [1-\cos \phi_1]}$. The exit velocity is then (remember $t_F=l/v_x$) 
\begin{equation}\label{eq:vnp1}
 v_2= v^{\rm isc}(\phi_1+ \omega l/v_x,\tilde v)= v_1 \frac{{\rm e}^{\frac{a}{h}[1-\cos (\phi_1+\omega l/v_x)]}}{{\rm e}^{\frac ah[1-\cos \phi_1]}}. 
\end{equation}
This procedure can be repeated again and again, yielding $v_n=\sum_{i=1}^n \triangle v_i + v_1$, where the $\triangle v_i$ are obtained in the above described manner by exploiting the piecewise motion on the invariant spanning curves. 
However, this sum is not suitable to obtain a closed expression for $v(t)$ 
(or $\langle |v|\rangle(t)$), especially since it contains the random phase shifts of $\pi$. We thus employ in the next section a random walk model, 
based on the statistical properties of the just described procedure,  and will thus be able to  calculate explicitly the exponential growth rate.  

\section{Random walk model}\label{ch:rwm}
The velocity in the random walk model is written as:
\begin{equation}
 v_{n+1}= v_n + \triangle v_n,
\end{equation}
where $\triangle v_n$ can be positive or negative. Since the  $\triangle v_n$ are determined by moving along parts of the invariant spanning curves of the FUM, and for the latter we know that $\triangle v^{\rm isc}\sim \tilde v$, we conclude that the $\triangle v_n$ are proportional to $v_n$, so $\triangle v_n=\pm c v_n $, yielding
\begin{equation}\label{rwmod}
 v_{n+1}=v_n\pm c\cdot v_n = (1\pm c) v_n
\end{equation}
The constant $c$ is an effective constant, besides the geometry of the billiard, $c$ depends in particular on the entry phase $\phi$ of a particle into the FUM and of course on the time $t_F=l/v_x$ 
(or phase $\triangle \phi = \omega l/v_x \mod 2\pi$) the particle spends in the FUM. We consider an ensemble of particles all starting with the same $v_x$, thus the phase shift $\triangle \phi$ is the same for all particles and $\triangle \phi$ is a constant. 
Under these assumptions we will proceed as follows: $\bullet$~We show that the random walk model of Eq.~\eqref{rwmod} leads to exponential Fermi acceleration.
$\bullet$~We will determine the effective $c$ using the phase space properties of the FUM.
To this end, we consider a particle after $N$ steps, i.e. $N$ cycles through the FUM. The probability $p(k)$ to have completed $k$ more positive than negative steps $\triangle v$ is then
\begin{equation}
p(k)=\frac{1}{2^N}\binom{N}{(N+k)/2}.
\end{equation}
We define $v_{(k)}$ as the velocity that is reached after $N$ steps with $k$ steps more in positive than in negative direction and $v_{(-k)}$ as the velocity that is reached after $N$ 
steps with $k$ steps more in negative than in positive direction.  There are of course many different paths, leading to the same $v_{(k)}$, however the order of the steps is irrelevant. 
Let us assume (without loss of generality) $N$ to be even, then $k$ can be any even number between $-N$ and $N$. The expectation value of the modulus of the velocity after $N$ 
steps is given by summing over all possible $k$'s and weighting them with the corresponding probability $p(k)$: 
\begin{eqnarray}\label{eq:rwm}
\left \langle |v|\right \rangle_N &=& \sum\limits_{k=-N,\;k\;\text{even}}^{N}  p(k) |v|_{(k)} \nonumber \\ &=&
\frac{1}{2^N}\sum\limits_{k=-N/2}^{N/2} \binom{N}{N/2+k} |v|_{(2k)}
\end{eqnarray}
By setting $\gamma:=1+c$ and $v_0$ as the initial velocity, using Eq.~\eqref{rwmod} for the positive sign, we obtain
\begin{equation}
\left \langle |v|\right \rangle_N=v_0 \frac{\left(1+\gamma^2\right)^N}{\left(2\gamma\right)^N}.
\end{equation}
To switch from the number of cycles $N$ to the actual time $t$, we use that the time between two collisions with the vertical walls 
is given by $L/v_{x}$. Substituting $N=v_{x} t/L$ yields the exponential time-law
\begin{equation}
\left \langle |v|\right \rangle (t)=v_0 {\rm e}^{R \cdot t}
\end{equation}
where the growth rate $R$ is given by
\begin{equation}\label{eq:growth_rate}
 R=\frac{v_{x}}{L}\ln\left(\gamma/2+1/2\gamma\right).
\end{equation}
\begin{figure}
\includegraphics[width=8.6cm]{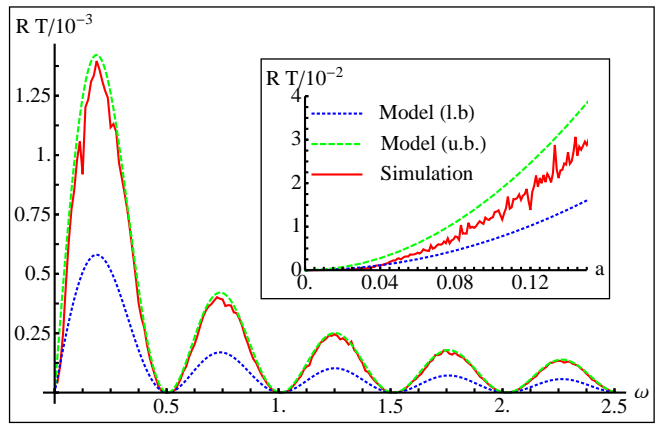}
\caption{(Color online) Frequency dependence of the growth rate $R$. The red line shows the result from the numerical simulations and the blue line the growth rate as predicted by our random walk model, 
c.f. Eqs.~\eqref{eq:growth_rate} and \eqref{eq:gamma_eff} ($a=0.025$). The inset shows the amplitude dependence ($\omega=0.1$) 
of the growth rate.  \label{fig:fig5}}
\end{figure}
The random walk model reproduces an exponential dependence of the ensemble averaged velocity on time. 
Nevertheless, in order to determine the system-specific value of the growth rate $R$,
we still have to determine the effective constant $c$. To this end, we rewrite the part of Eq.~\eqref{rwmod} with the `+' as $c=(v_{n+1}-v_n)/v_n$. The velocity $v_{n+1}$ is, according to Eq.~\eqref{eq:vnp1}, given by 
\begin{equation}
 v_{n+1}= v^{\rm isc}(\phi_n+ \triangle \phi,\tilde v(v_n,\phi_n)),
\end{equation}
where we write $\tilde v = \tilde v(v_n,\phi_n)$, since $\tilde v$ depends on the entry phase $\phi_n$ and the entry velocity $v_n$. 
For a fixed $\triangle \phi$, we thus have $c=c(v_n,\phi_n)=v^{\text{isc}}(v_n,\phi_n+\triangle \phi)/v^{\text{isc}}(v_n,\phi_n\phi) - 1$ with $v^{\text{isc}}\propto v_n$ leading to $c(v_n,\phi_n)=c(\phi_n)$. 
Thus, the effective $c$ is given by averaging over all entry phases $\phi_n$ that lead to a positive $\triangle v_n$ 
(This is sufficient, since the random walk model of Eq.~\eqref{rwmod},\eqref{eq:rwm} has intrinsically included the `-' part, 
allowing the $\triangle v$ to be negative.):
\begin{equation}
 c_{\rm eff}= \frac{1}{\phi_{n,2}-\phi_{n,1}}\int\limits_{\phi_{n,1}}^{\phi_{n,2}}  c(\phi_n) {\rm d}\phi_n, 
\end{equation}
where
\begin{equation} \label{eq:effc}
 c(\phi_n)= \frac{{\rm e}^{\frac{a}{h}[1-\cos (\phi_n+\triangle \phi)]}}{{\rm e}^{\frac{a}{h}[1-\cos \phi_n]}}-1.
\end{equation}
The integral over $\phi_n$ has to be evaluated such that $\triangle v_n$ is positive in order to account all accelerating trajectories, 
i.e. $\phi_{n,1}=-\triangle \phi/(2 \omega)$ and $\phi_{n,2}=\pi/\omega-\triangle \phi/(2 \omega)$ yielding for the normalization $N_{\phi}= 1/(\phi_{n,2}-\phi_{n,1})=\omega/\pi$. 
Since the growth rate $R$ depends on $\gamma$, see Eq.~\eqref{eq:growth_rate}, and $\gamma = 1+c$, we finally obtain
\begin{eqnarray}\label{eq:gamma_eff}
 \gamma_{\rm eff} &=& 1+c_{\rm eff}=\frac{\omega}{\pi}\int\limits_{-\Delta \phi/2\omega}^{\pi/\omega-\Delta\phi/2\omega} \frac{{\rm e}^{\frac{a}{h}[1-\cos (\phi_n+\triangle \phi)]}}{{\rm e}^{\frac{a}{h}[1-\cos \phi_n]}} {\rm d}\phi_n
\nonumber \\ &=:& 1+ \langle c \rangle_\phi.
\end{eqnarray}
Here we assumed that the entry phases $\phi_n$ are uniformly distributed, which is valid on a sufficient long time scale if the ratio $L/v_x$ is incommensurate with the driving period $T=2\pi/\omega$. 
Note that the growth rate $R$ of Eq.~\eqref{eq:growth_rate} together with the corresponding $\gamma$ of Eq.~\eqref{eq:gamma_eff} has been obtained ab initio without any fit parameters. 
However, the corresponding values are too small. From the results of the simulation shown in Fig.~\ref{grate}, for $a=0.1$ and $\omega=0.02$ we obtain for the growth rate $R \approx 1.1\cdot 10^{-5}$. 
Inserting Eq.~\eqref{eq:gamma_eff} into Eq.~\eqref{eq:growth_rate} leads to $R=0.51\cdot 10^{-5}$, which provides the order of magnitude of the numerical result, but is too small by about a factor 2.
The reason for this is as follows: The sum of two sequences of the dynamics each consisting of e.g. three steps, one with a $c(\phi)$ close to the maximally possible value $c_{\rm max}$ in each step and one 
with steps close to the minimal value of $c(\phi)$ (i.e. $c_{\rm min}=0$) contributes more significantly to the ensemble average $\langle |v|\rangle$, than the sum of two corresponding sequences both with step width
$c=(c_{\rm max}+c_{\rm min})/2$. However, according to Eq.~\eqref{eq:gamma_eff} we calculated a mean of the latter type and therefore obtained a lower bound of the correct growth rate.
The deviation of the result obtained from the simulation and the one from the model can thus 
be understood as a consequence of the negligence of correlations.  
One way of effectively including correlations to the definition of $c_{\rm eff}$ is to define $c_{\rm eff}= \langle c^m\rangle_\phi /\langle c^{m-1}\rangle_\phi$ with $m>1$.
Therein $\left\langle c^m\right \rangle_\phi$ denotes the average of $c^m(\phi)$ over all phases in the interval $[-\triangle\phi/(2\omega),(\pi-\triangle\phi/2)/\omega]$ with $c$ given by Eq.~\eqref{eq:effc}. 
An upper bound for the growth rate can be obtained 
within the assumption that all steps but those with maximal $c(\phi)$ are suppressed, i.e. by calculating the effective $c$ with the assumption that $c(\phi)$ is equal to the maximally possible
value $c_{\rm max}$ at each step:
\begin{eqnarray}\label{eq:gamma_max}
\gamma_{\rm max}&=& 1+ c_{\rm max} = \frac{{\rm e}^{\frac{a}{h}[1-\cos (\pi/2+\triangle \phi/2)]}}{{\rm e}^{\frac{a}{h}[1-\cos (\pi/2 - \triangle \phi/2)]}} \nonumber \\
& = & 1+\lim\limits^{}_{N\rightarrow \infty}\frac{\langle c^N\rangle_\phi}{\langle c^{N-1} \rangle_\phi}
\end{eqnarray}
This upper bound leads to $R=1.24\cdot 10^{-5}$, which is quite close to the result obtained from the simulation.
In order to test these estimations for a whole range of parameters and also to show that our random walk model correctly describes the 
whole dependency of the growth rate of the parameters of the system, we 
 extract the growth rate $R$ for different driving frequencies $\omega$ at fixed amplitude $a=0.025$ by performing a numerical simulation for each value of the 
frequency and compare $R$ with the corresponding result obtained from our random walk model. 
The growth rate $R(\omega)$ (see Fig.~\ref{fig:fig5}) shows characteristic (decaying) oscillations, 
as already theoretically predicted in Ref.~\cite{Shah:2010}. The minima where $R(\omega)$ is exactly zero can be easily understood. At these values of $\omega$, 
the driving period $T=2\pi/\omega$ and the time between two collisions with the same vertical wall $2L/v_x$ are commensurable, what leads
to a $\phi$-periodic entering and leaving of the FUM for $v_y \rightarrow \infty$ (when the first and the last collision with the oscillating bar are converged to its edges).
Note that the occurrence of the minima in the
growth rate $R(\omega)$ are based on the fact, that all particles of the ensemble possess the same, constant velocity in $x$-direction. Apparently, these characteristic oscillations are fully reproduced by our model.

The inset of Fig.~\ref{fig:fig5} shows the analogous comparison between model and simulation
for a fixed driving frequency $\omega=0.1$ and different but small values of the amplitude (the regime where the invariant spanning curves of the 
FUM can be well approximated within the SWA). From this, we firstly observe that the growth rate strongly increases with the amplitude of the oscillating bar, and secondly, that 
this dependency can be well explained by the random walk model. 
According to the good agreement between the simulation and the model, we may conclude from
Eq.~\eqref{eq:gamma_eff} and Eq.~\eqref{eq:growth_rate} that the amplitude-dependence of the growth rate is approximately given by $R(a)\propto \ln\left(\cosh(a/h)\right)$.
As Fig.~\ref{fumiscs} reveals, we may not expect that this is also true for large values of the amplitude, since then 
the expressions for the invariant spanning curves obtained within the static wall approximation strongly differ from the numerical results.
Obviously, the result of the simulation is between the estimations for the lower and the upper bound for all values of the system parameters.
\\These results indicate that all details of the specific system under consideration which are not accounted for the random walk model including the existence of a chaotic sea do not contribute crucially to the growth rate.
Even more, all details of the specific system are only needed to calculate the effective $c$. 
The requirements for the occurrence of exponential Fermi acceleration are comparatively weak: A temporally periodic entering and leaving of invariant spanning curves
with $v^{\rm isc}_{\rm max}-v^{\rm isc}_{\rm min} \propto v^{\rm isc}_{\rm min}$ at different phases is sufficient.

\section{Conclusion}\label{ch:conclusion}
In this work, we have investigated the physical mechanism leading to exponential Fermi acceleration (FA) 
in the rectangular billiard with an oscillating bar inside. In particular, 
we showed that the dynamics of individual trajectories can be understood as alternating phases 
of motion in an appropriate one-dimensional Fermi-Ulam model (FUM) and free propagation.  
During the temporal FUM phases, the particles move (in the high velocity regime which is of interest here) 
on invariant spanning curves of the FUM, which can be - at least for small driving amplitudes - 
obtained analytically within the static wall approximation. Using the intrinsic property of the invariant spanning curves of the 
FUM that the difference between the maximal and the minimal velocity grows linearly 
with the minimal velocity of the invariant curve and the fact that acceleration and deceleration have equal probability, the 
process can be modeled as a random walk with step width proportional to the velocity itself, i.e. $v_{n+1}=v_n\pm c v_n$. 
This model explains the occurrence of the exponential acceleration.
Calculating an effective step width $c_{\rm eff}$ we obtain a good ab initio estimation of the growth rate and reproduce the whole 
qualitative dependency of the system parameters.
We emphasize, that our random walk model reflects that a temporally periodic entering and leaving of equally shaped invariant spanning curves
which have the property that the difference between the maximal and the minimal velocity grows linearly 
with the minimal velocity of the latter are the key ingredients for the occurrence of exponential Fermi acceleration (all details of the specific system are contained in the factor $c$).
This opens the perspective of searching for other systems 
exhibiting the phenomenon of exponential Fermi acceleration.

\section{Acknowledgments} 
Financial support by the DAAD in the framework of an exchange program with Greece (IKYDA) is acknowledged.
P.S. gratefully appreciates financial support by the Deutsche Forschungsgemeinschaft.

\end{document}